\documentclass[twocolumn,nofootinbib,superscriptaddress]{revtex4-1}

\pdfoutput=1

\usepackage{slashed}

\usepackage[normalem]{ulem}

\usepackage{color}

\usepackage{epsfig}
\usepackage{epstopdf}
\usepackage{epsf}
\input{epsf.sty}
\usepackage{graphicx,amsmath,amssymb}
\usepackage{epsfig}
\allowdisplaybreaks

\def\ei{\end{itemize}}
\def\be{\begin{equation}}
\def\ee{\end{equation}}
\newcommand{\bea}{\begin{eqnarray}}
\newcommand{\eea}{\end{eqnarray}}

\usepackage{bbm,bm,amsmath,amssymb}

\usepackage{hyperref}

\usepackage{float}

\begin{document}

 \title{\Large\rm {\bf   \boldmath  B-mode  Targets}}

\author{Renata Kallosh}
\email{kallosh@stanford.edu}
\author{Andrei Linde}
 \email{alinde@stanford.edu}
\affiliation{Stanford Institute for Theoretical Physics and Department of Physics, Stanford University, Stanford,
CA 94305, USA}

 \begin{abstract}
The CMB-S4 Science Book \cite{Abazajian:2016yjj}, Simons Observatory report \cite{Ade:2018sbj}, Astro2020  Science White Paper on gravitational waves  \cite{Shandera:2019ufi}, and PICO report  \cite{Hanany:2019lle} contain an extensive discussion of many ongoing and planned efforts in the search for gravitational waves produced by inflation. Here we give a short executive summary of the results obtained in  our  papers \cite{Kallosh:2018zsi,Kallosh:2019jnl,RKAL1},  which specify the simplest available inflationary models providing physically motivated targets for these searches. Our conclusions are specific for the  $10^{-3} \lesssim r \lesssim 10^{-2}$ range, where we present  the B-mode benchmarks of the  U-duality symmetric  class of $\alpha$-attractors, and for $ r \lesssim 10^{-3}$, where we present B-mode targets, for which the future precision measurements of $n_s$  will be decisive. We show that a combination of the simplest $\alpha$-attractors and KKLTI models of D-brane inflation covers most of the area favored by Planck 2018.

 \end{abstract}

\maketitle


\parskip 2 pt  
\section{Introduction}

The goals of  future CMB missions can be illustrated by   Fig. 2.2  in the PICO report \cite{Hanany:2019lle}, reproduced here:
\begin{figure}[!h]
\hskip -10pt
\includegraphics[width= 9cm]{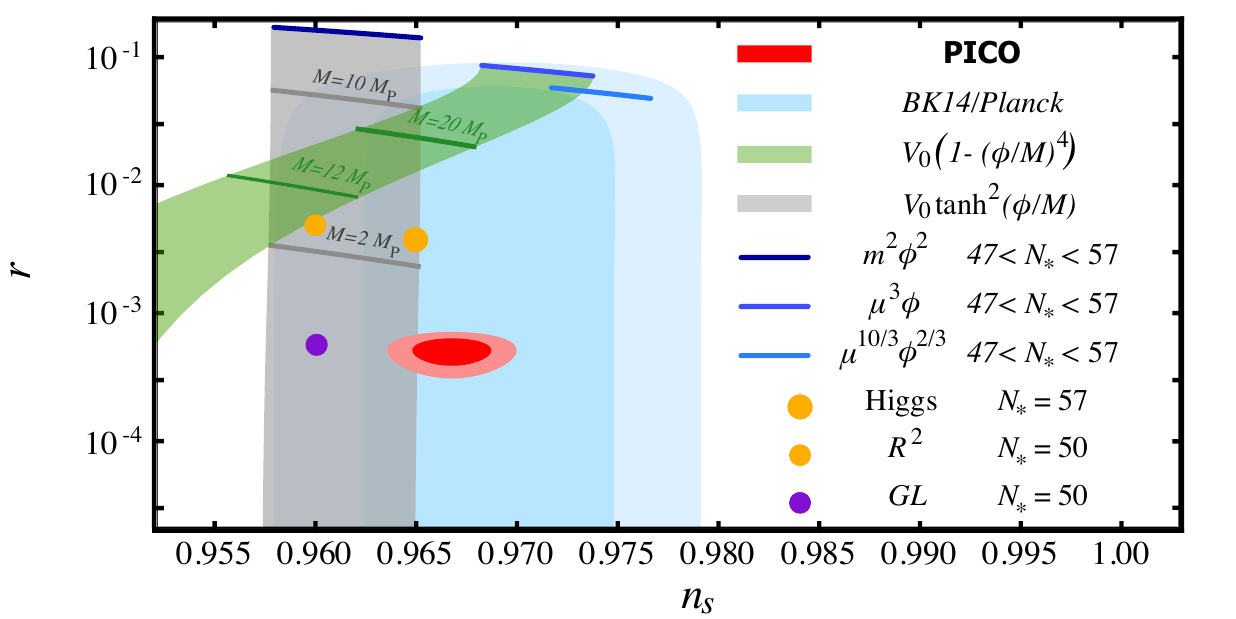}
\caption{\footnotesize This is    Fig. 2.2 from the  PICO report \cite{Hanany:2019lle}. The blue region is based on the results of BICEP/Keck 14 and Planck 2015.}
\label{WhitePICO}
\vspace{-0.1cm}
\end{figure}

\noindent PICO plans to detect primordial gravitational waves if  $r\gtrsim 5\times 10^{-4} (5\sigma)$. It is equally important that the constraints on $n_s$ can be significantly improved, to achieve $\sigma (n_s)\sim 0.002 - 0.0015$ \cite{Ade:2018sbj,Hanany:2019lle}.

In this paper we will discuss the simplest inflationary models which describe all presently available data, and  identify some `future-safe' models, which have a fighting chance to describe all data on  spectral index $n_{s}$ and tensor to scalar ratio $r$ to be obtained in the next one or two decades. 
There are some very interesting targets in the range $r \sim 4 \times 10^{-2}$, such as the string theory related  axion monodromy models \cite{Silverstein:2008sg,McAllister:2008hb,McAllister:2014mpa}. If we discover inflationary B-modes in this range, it will be fantastic, but what if we go through the range $r \gtrsim 10^{{-2}}$ without finding the signal? Do we have any other legitimate targets for the future missions discussed in \cite{Abazajian:2016yjj,Shandera:2019ufi,Hanany:2019lle}?

The results of our investigation of these issues is contained in \cite{Linde:2016hbb} and in our  papers \cite{Kallosh:2018zsi,Kallosh:2019jnl,RKAL1}. Our analysis was focused on the models favored by the recent Planck 2018 results, shown in Fig.~8 and Table~5 of \cite{Akrami:2018odb}, and by BICEP2/Keck 2014 results shown by  Fig.~5 in \cite{Ade:2018gkx}.  These models include $\alpha$-attractors \cite{Kallosh:2013yoa,Carrasco:2015uma,Kallosh:2017wnt}, providing a significant generalization of the Starobinsky model  \cite{Starobinsky:1980te} and the Higgs inflation model \cite{Salopek:1988qh,Bezrukov:2007ep}. We also discuss   hilltop inflation  \cite{Linde:1981mu,Boubekeur:2005zm}, and   D-brane inflation  \cite{Dvali:2001fw,Burgess:2001fx,GarciaBellido:2001ky,Kachru:2003sx,Martin:2013tda}. Predictions of the simplest $\alpha$-attractor models are shown by the vertical yellow stripe in Fig. 8 of  \cite{Akrami:2018odb} and by the grey band in  Fig. 2.2 from  PICO, which is  our Fig.~\ref{WhitePICO}. Predictions of the simplest hilltop models with the potential $V\sim 1-{\phi^{4}\over m^{4}}$ are shown by the green area in these figures. 

Since our papers  \cite{Kallosh:2018zsi,Kallosh:2019jnl,RKAL1} are large and rather technical, targeted for the hep-th audience,  we decided to give here a short executive summary of our main conclusions.

We have  found in \cite{Kallosh:2019jnl} that for $ m > 10 M_{P} $, which is the only regime where the hilltop models  $V \sim  1-{\phi^4\over m^4 }$ could be data-compatible, their predictions  are directly related to the unboundedness of their potential from below, which leads to collapse of the universe soon after the end of  inflation. Improved versions of these models typically have very different model-dependent predictions, unrelated to inflation at the hilltop, with significantly higher values of $r$.  Therefore  we  removed the green area corresponding to  the hilltop model $V \sim  1-{\phi^4\over m^4 }$  from  Figs. \ref{stripes}, \ref{branes} and \ref{bluebranes} illustrating our results to be presented below. 

The simplest D-brane inflation models with  $V \sim  1-{m^n\over \phi^n}$ are also unbounded from below; they are called BI models in   the `Encyclop\ae dia Inflationaris' \cite{Martin:2013tda}. We  eliminated them  from our figures since they are not physically consistent. But at small $m$,  their predictions nearly coincide with the ones of the improved D-brane inflation models $V = {\phi^{n}\over \phi^{n}+ m^{n}}$,  which were called KKLTI in \cite{Martin:2013tda}. We will describe these models below.

The predictions of the simplest $\alpha$-attractors and KKLTI models  are well inside the $2\sigma$ region for $n_s$ and $r$ in the Planck 2018 data.  Following the way of evaluation of inflationary models used in the Planck 2018 data release, in Figs. \ref{stripes} and \ref{branes} we make a comparison using the CMB data only. The dark (light) pink areas in the these figures correspond to $n_s$ and $r$ favored by  CMB-related data at the 1$\sigma$ (2$\sigma$) level. Inclusion of the BAO   leads to very similar results, see   Fig. \ref{bluebranes} and \cite{Kallosh:2018zsi,Kallosh:2019jnl,RKAL1}.

\section{U-duality benchmarks for  $10^{-3} \lesssim r \lesssim 10^{-2}$}
  \begin{figure}[!h]
\vspace*{-3mm}
\begin{center}
\hskip -5mm\includegraphics[width=7.2cm]{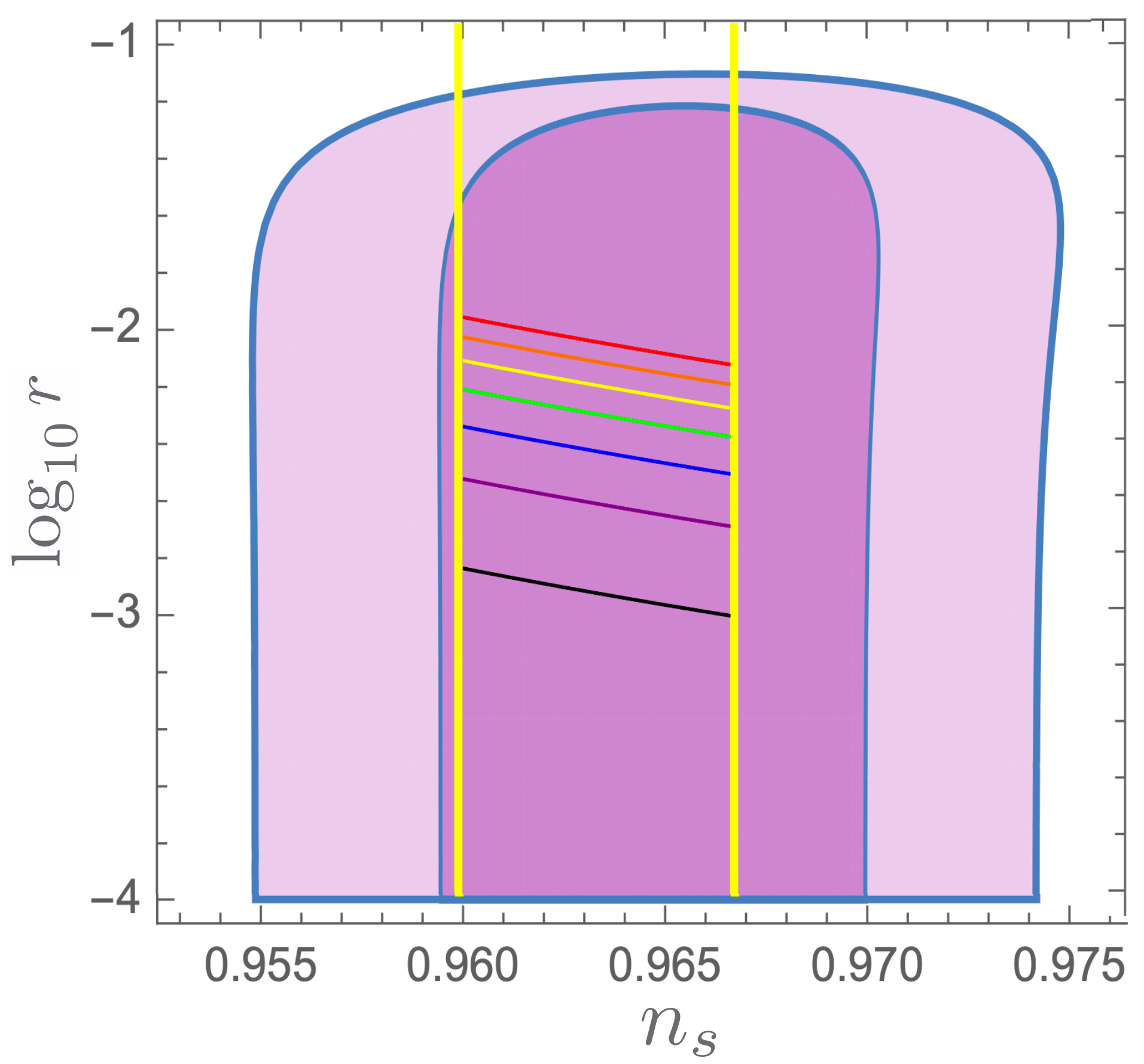}\vspace{-6pt}
\caption{\footnotesize  The inflationary models between two yellow stripes are the  simplest $\alpha$-attractor models with $V \sim \tanh^{2}{\phi\over \sqrt {6\alpha}}$ for  $ 50< N<60$.  In the range $10^{-3} \lesssim r \lesssim 10^{-2}$ there are U-duality benchmarks associated with  M-theory, string theory, maximal ${\cal N}=8$  supergravity. They take seven
 discrete values  $3\alpha=7,6,5,4,3,2,1$,  from $3\alpha=7$ in red all the way to    $3\alpha=1$ in black. The intermediate blue line with $\alpha=1$ represents also  Starobinsky  and Higgs inflationary models. }
\label{stripes}
\end{center}
\end{figure} 


In Fig. \ref{stripes} we present several discrete benchmarks for  $\alpha$-attractor models, in units $M_{P} = 1$. 
 The $n_s$ and $r$ predictions of  these models in the small $\alpha$ limit depend only on their geometric kinetic terms.  
 The benchmark models have kinetic terms originating from maximal supersymmetry (M-theory, string theory, maximal ${\cal N}=8$ supergravity), spontaneously broken to the minimal ${\cal N}=1$ supergravity. The 7-disk model \cite{Ferrara:2016fwe,Kallosh:2017ced} allows 7  discrete values $3\alpha=7,6,5,4,3,2,1$.
 The characteristic scale $M$ of the inflationary potential for $\alpha$-attractors  is $\sqrt{3\alpha/ 2}$ \cite{Linde:2016hbb}. The last maximal supersymmetry benchmark has $3\alpha=1$, which at   $N=60$ corresponds to    $r\approx 10^{-3}$.
 
At $3\alpha = 6$ we would probe string theory fibre inflation \cite{Cicoli:2008gp,Kallosh:2017wku}, at $3\alpha =3$ we would probe the Starobinsky model~\cite{Starobinsky:1980te}, the Higgs inflationary model~\cite{Salopek:1988qh,Bezrukov:2007ep},  as well as the conformal inflation model~\cite{Kallosh:2013hoa}.  Finally, at $3\alpha =1$ we would probe the case of the maximal superconformal symmetry, as explained in  \cite{RKAL1}.

 There is yet another target, at $\alpha = 1/9$, $r \sim 5\times 10^{{-4}}$, which corresponds to the GL model \cite{Goncharov:1985yu,Linde:2014hfa} shown by a purple dot in the PICO Fig. \ref{WhitePICO}. This is a supergravity inflationary model  involving just a single superfield, providing the first example of chaotic inflation with a plateau potential. There are some mathematical reasons for the specific value of $\alpha = 1/9$ in this model, but we believe that the targets at $3\alpha=7,6,5,4,3,2,1$ are better motivated from the point of view of fundamental physics.
 
 Independently of these specific targets, there is a certain aspect of all $\alpha$-attractors that deserves particular attention. The value of $\alpha$, which can be found by measurement of $r$, is directly related to the curvature of the moduli space \cite{Ferrara:2013rsa,Kallosh:2015zsa}:
$ -{\cal R}={2\over 3 \alpha} = M^{{-2}} = {8\over r\, N^{2}}$, 
where $M$ is the characteristic scale of inflation for $\alpha$-attractors \cite{Abazajian:2016yjj,Linde:2016hbb}.   Thus, investigation of the gravitational waves produced during inflation may go beyond investigation of our space-time: It may help us to study geometry of the internal space of scalar fields responsible for inflation.

\section{ Attractor stripes at $ r \lesssim 10^{-3}$   }

\begin{figure}[!h]
\begin{center}
\hskip 0.4cm\includegraphics[width=6.3cm]{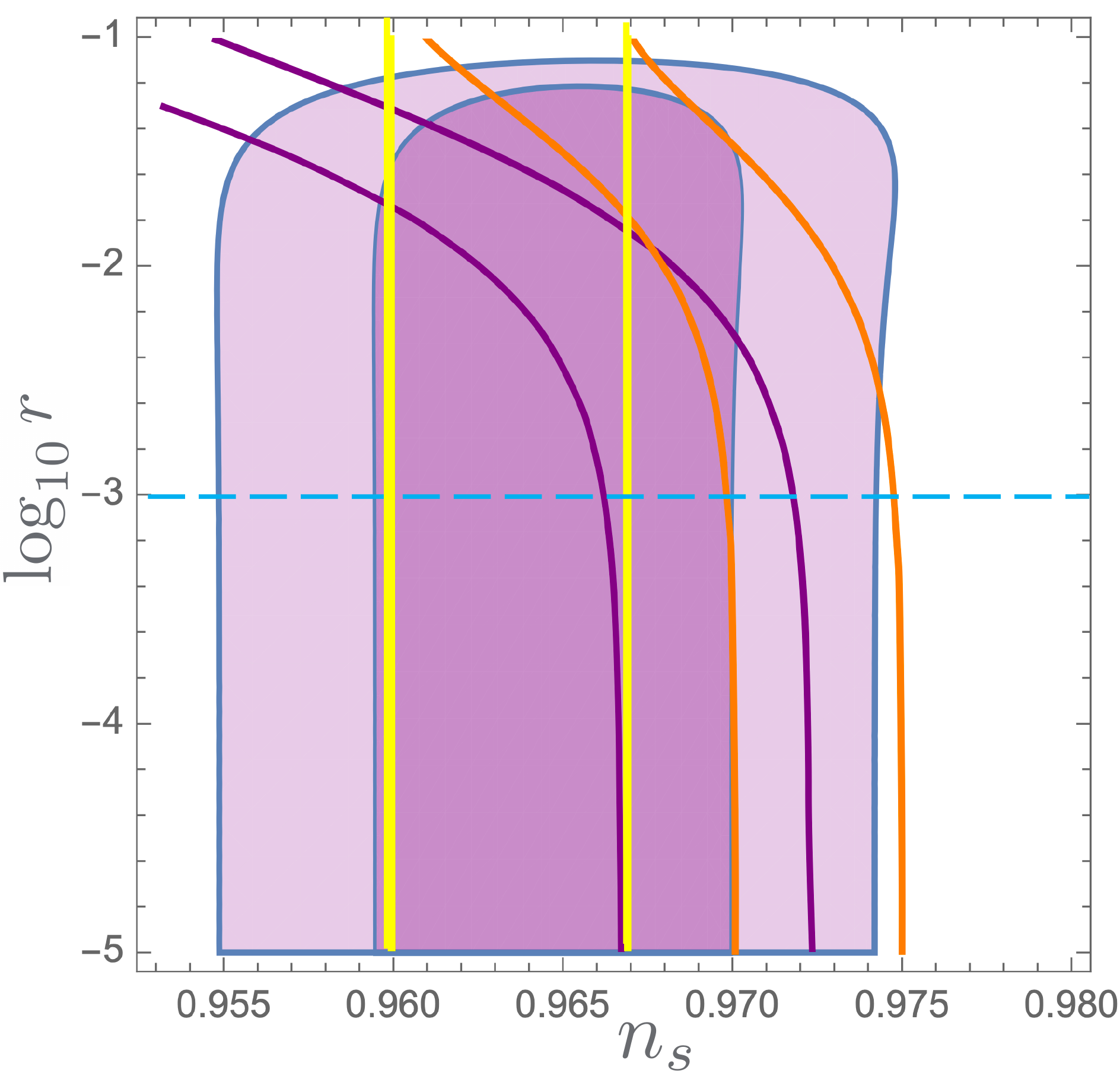}
\vskip -0.2cm\caption{\footnotesize Predictions of $\alpha$-attractors  and KKLTI models.  Two yellow lines  are for the simplest $\alpha$-attractor models for $N = 50$ and $N = 60$.  Two purple lines are for the  KKLTI D3-brane model with $n = 4$. Two orange lines  are for the  KKLTI D5-brane model with $n = 2$.  For $ r \lesssim 10^{-3}$, indicated by the blue dashed line, the  predictions of all models converge to their attractor values shown by the vertical lines. }
\label{branes}
\end{center}
\end{figure}

\begin{figure}[!h]
 \vspace*{3mm}
\begin{center}
\includegraphics[width=6.5cm]{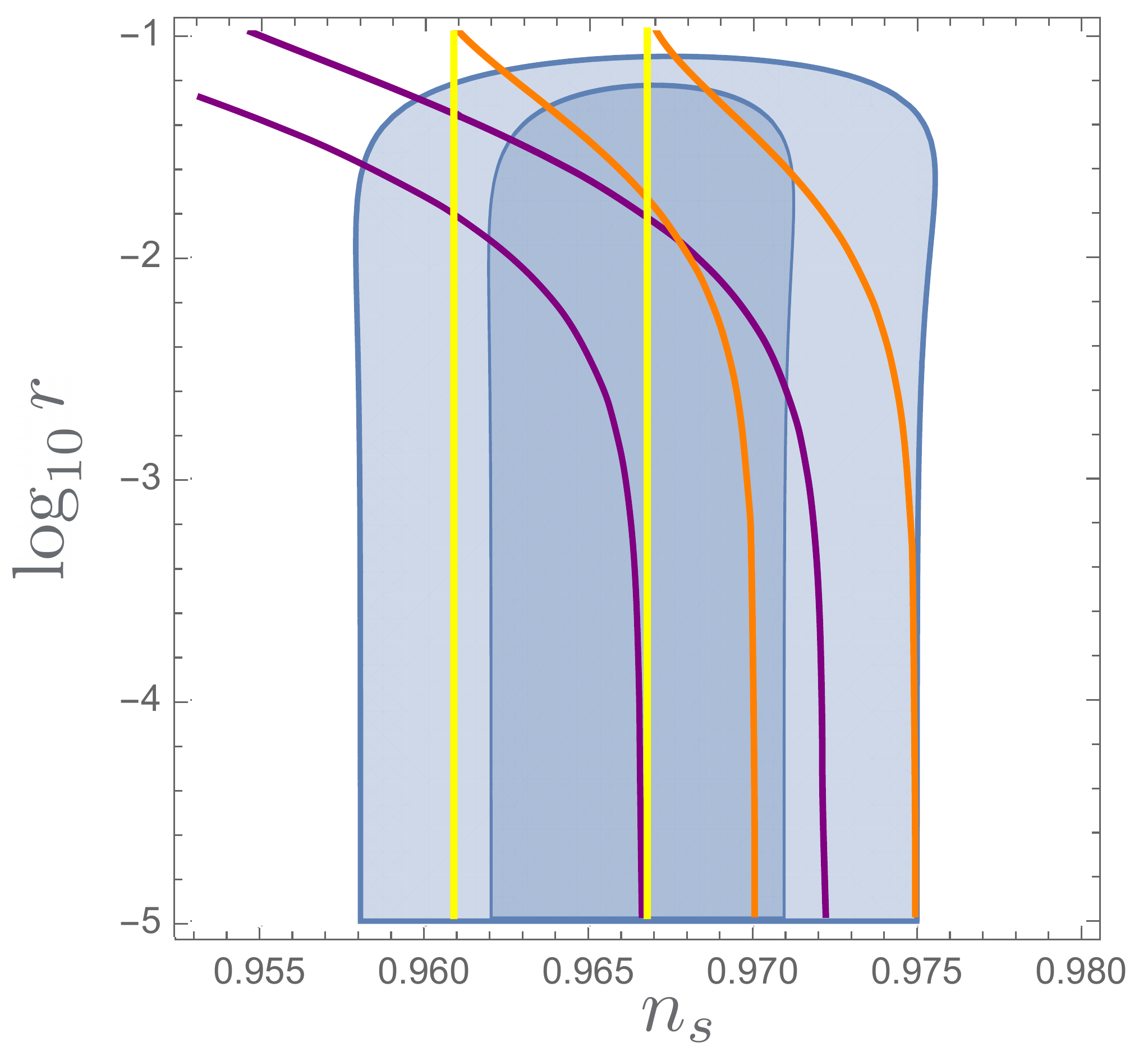}
\vskip -0.2cm\caption{\footnotesize For completeness, we plot here predictions of $\alpha$-attractors  and KKLTI models in comparison with the Planck 2018, including BAO.  As one can see from this  figure and the previous one, a combination of the simplest $\alpha$-attractors and KKLTI models of D-brane inflation covers most of the area favored by Planck 2018. Notice the shrinking of the range of $n_{s}$ in the blue area for Planck 2018 as compared with the Planck 2015 results shown in Fig. \ref{WhitePICO}.}
\label{bluebranes}
\end{center}
\end{figure}

KKLTI  potentials described in \cite{Kachru:2003sx,Martin:2013tda} are $V =V_{0} {\phi^{n}\over \phi^{n}+ m^{n}}$. They have a minimum at $\phi = 0$ and a plateau $V = V_{0}$ at large $\phi$. Recently these models were re-examined in \cite{Kallosh:2018zsi,Kallosh:2019jnl}. The main results are shown in Figs.~\ref{branes} and \ref{bluebranes}.

\vskip -0.5cm
\vskip -0.2cm

These figures show  that a combination of the simplest $\alpha$-attractor model $V \sim \tanh^{2}{\phi\over \sqrt {6\alpha}}$ with the two simplest KKLTI models almost completely covers the area in the ($n_{s},r$) space favored by Planck 2018. At $r\lesssim 10^{{-3}}$, $\alpha$-attractors predict $1-n_{s }= {2\over N} $, KKLTI models   with $n = 4$ predict   $1-n_{s }= {5\over 3N}$, and KKLTI models with $n = 2$ predict  $1-n_{s }= {3\over 2N}$.  This  addresses the concerns expressed in \cite{Shandera:2019ufi} that it could be difficult to find   inflationary models with $r \lesssim 10^{{-3}}$ and $1-n_{s} \sim c/N$  with $c = O(1)$. As one can see, such models are already known. They are prominently represented in the Planck 2015 and Planck 2018 data releases  \cite{Akrami:2018odb,Ade:2015lrj}.

We are unaware of any important discrete targets for  $ r\ll 10^{-3}$, with the possible exception of the GL model. In some string theory models one may have $r$  as small as $r\sim 10^{-6} -10^{-10}$   \cite{Kachru:2003sx,Martin:2013tda}. However, the expected significant improvement of the constraints on $n_{s}$ by  2 or 3 times relative to Planck 2018 
 \cite{Ade:2018sbj,Hanany:2019lle} would allow to discriminate between different  models shown in Fig.~\ref{branes}. At present, the yellow, the purple and  the orange attractor stripes fit the data, but a more precise knowledge  of $n_s$ may help us to distinguish these models, and learn more about the post-inflationary evolution in these models, including reheating, which affects the required values of N, and, consequently, $n_{s}$. Therefore at  $ r \lesssim 10^{-3}$, where the predictions of $\alpha$-attractors and KKLTI models do not depend on $r$, we have interesting targets for $n_s$. 

\begin{figure}[!h]
\begin{center}
\hskip -0.3cm
\includegraphics[width=8.5cm]{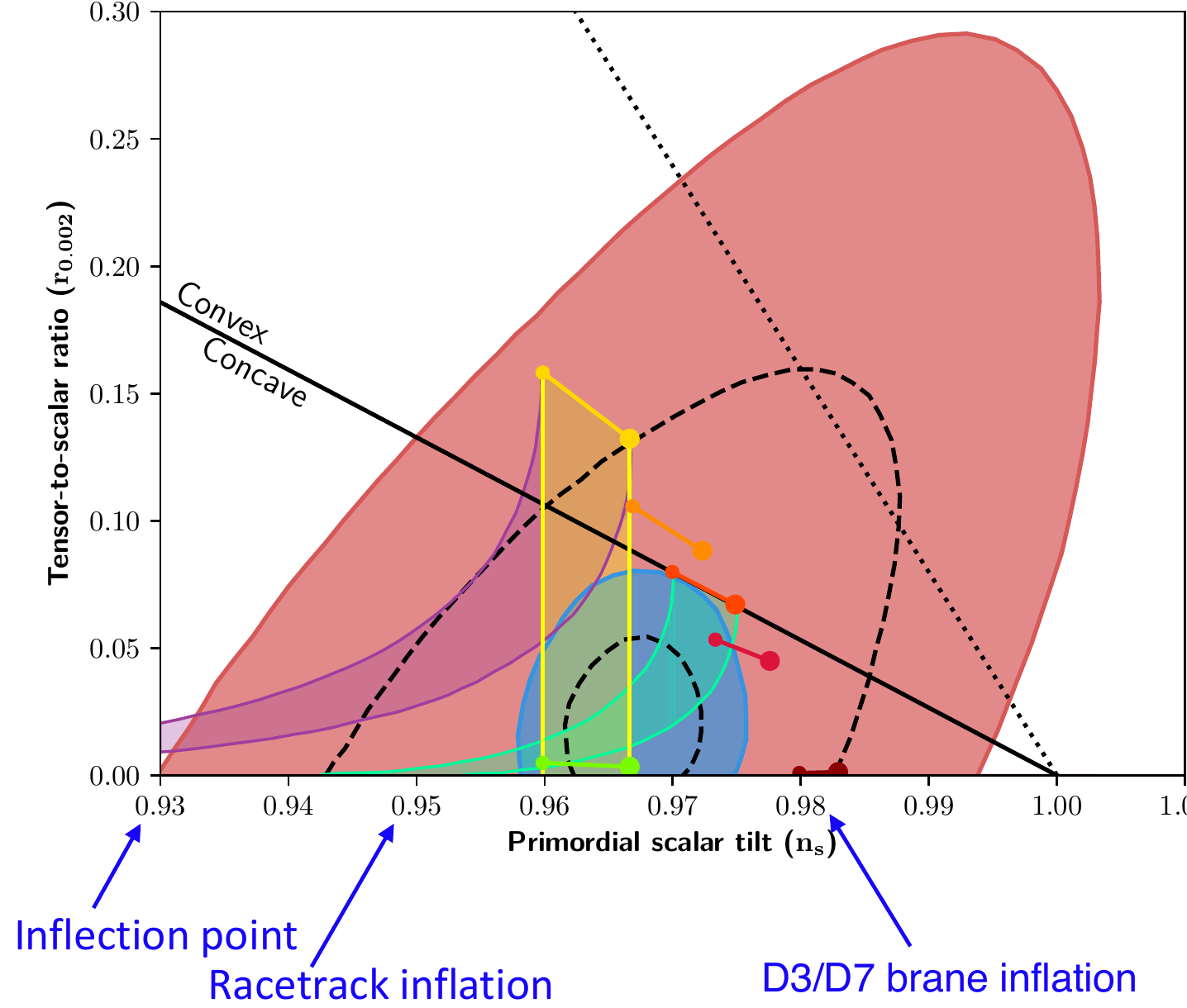}
\caption{\footnotesize  Many favorite string inflation models from a decade ago, with very low $r$,  are now
ruled out by precision data on $n_s$.  7-year WMAP results \cite{Komatsu:2010fb} are in red, Planck 2018 results  \cite{Akrami:2018odb} are in blue.}
\label{ns}
\end{center}
\end{figure}

 We have had a very similar experience during the last decade. The improvement of the precision of determination of $n_s$ during the 8 years between  the release  of WMAP data  in 2010 \cite{Komatsu:2010fb} to  Planck 2018 \cite{Akrami:2018odb} shown in Fig. \ref{ns} from \cite{Akrami:2018ylq} was the main reason for elimination of many inflationary models predicting extremely low $r$.

\section{Conclusions}
This paper gives a brief summary of the results obtained in our  papers  \cite{Kallosh:2018zsi,Kallosh:2019jnl,RKAL1}, which may have some implications for planning of the future B-mode searches. 
In particular, we found that the predictions of the simplest hilltop inflation models shown by the green area in Fig. \ref{WhitePICO} do not constitute legitimate targets for B-mode searches, because these predictions are directly related to the physical inconsistency of these models \cite{Kallosh:2019jnl}. 

There is a popular but not universally applicable way to parametrize inflationary models by the assumption that they should satisfy  the relation $1-n_{s} = {p+1\over N}$, where $p = O(1)$ is a phenomenological parameter. This parametrization works  well for $\alpha$-attractors, but, as one can infer from Figs. \ref{branes}, \ref{bluebranes}, it applies to the KKLTI models  only for $r \lesssim 10^{{-3}}$,  at the lower boundary of the range of $r$ to be studied by CMB-S4.

On the other hand, we have found that $\alpha$-attractors, in combination with the KKLTI models, almost completely cover the dark   $1\sigma$ region in the $(n_{s}, r)$ space  favored by Planck~2018, all the way down to $r = 0$, as shown in Figs.~\ref{branes}, \ref{bluebranes}. Therefore, in addition to (and independently of) being the candidates for the role of a consistent inflationary theory compatible with the available observational data, these models can  provide a simple physically motivated parametrization of the future CMB results for the range of $n_{s}$ and $r$ favored by Planck 2018.

Investigation of these models  provides a list of specific B-mode  targets for the future B-mode searches. Seven different targets in the range $10^{-3}\lesssim r \lesssim 10^{-2}$ are shown in Fig.~\ref{stripes}, corresponding to discrete benchmarks representing  a set of U-duality invariant $\alpha$-attractor inflationary models.  An important aspect of the search of the gravitational waves in this context is that the knowledge of $r$ allows to find the curvature of the internal space of scalar fields responsible for inflation.

\vskip 0.1cm

 We are grateful to Z. Ahmed, F. Finelli, R. Flauger, M. Hazumi, S. Kachru, L. Knox, Chao-Lin Kuo, J. Martin, L. Page, D. Roest, L. Senatore, E. Silverstein, V. Venin, R. Wechsler and Y. Yamada for the stimulating discussions. This work is supported by SITP and by the US National Science Foundation grant PHY-1720397 and by the Simons Foundation Origins of the Universe program. 


\bibliography{lindekalloshrefs}

\end{document}